\documentclass[lettersize,journal]{IEEEtran}
\usepackage{amsmath,amsfonts}
\usepackage{algorithmic}
\usepackage{algorithm}
\usepackage{array}
\usepackage[caption=false,font=normalsize,labelfont=sf,textfont=sf]{subfig}
\usepackage{textcomp}
\usepackage{stfloats}
\usepackage{url}
\usepackage{verbatim}
\usepackage{graphicx}
\usepackage{cite}
\usepackage{bbm,mathrsfs,enumerate,latexsym,psfrag,amssymb}
\usepackage{epsfig,epstopdf,verbatim,color,multirow,booktabs}
\usepackage{flushend}
\makeatletter

\newcommand{\Rmnum}[1]{\expandafter\@slowromancap\romannumeral #1@}
\makeatother

\newtheorem{rem}{Remark}
\begin{document}

\title{Reconstructing Visual Stimulus Images from EEG Signals Based on Deep Visual Representation Model}

\author{Hongguang Pan$^*$, Zhuoyi Li, Yunpeng Fu, Xuebin Qin, Jianchen Hu
\thanks{This work is supported by the National Natural Science Foundation of China [51905416, 51804249], Xi'an Science and Technology Program [2022JH-RGZN-0041], Qin Chuangyuan ``Scientists + Engineers'' Team Construction in Shaanxi Province [2022KXJ-38], Natural Science Basic Research Program of Shaanxi [2021JQ-574], Scientific Research Plan Projects of Shaanxi Education Department [20JK0758]. \emph{($^*$Corresponding author: Hongguang Pan.)}}
\thanks{Hongguang Pan is with the College of Electrical and Control Engineering, Xi'an University of Science and Technology, Xi'an 710054, China, and also with the Xi’an Key Laboratory of Electrical Equipment Condition Monitoring and Power Supply Security, Xi’an 710054, China (e-mail: hongguangpan@163.com).}
\thanks{Zhuoyi Li, Yunpeng Fu and Xuebin Qin are with the College of Electrical and Control Engineering, Xi'an University of Science and Technology, Xi'an 710054, China (e-mail: zhuoyilee@163.com, yunpengfu@163.com and qinxb@xust.edu.cn).}
\thanks{Jianchen Hu is with the School of Automation Science and Engineering, Xi'an Jiaotong University, Xi'an 710049, China (e-mail: horace89@xjtu.edu.cn).}}

\markboth{IEEE Transactions on Human-Machine Systems}
{PAN \MakeLowercase{\textit{et al.}}: Reconstructing Visual Stimulus Images from EEG Signals}


\maketitle

\begin{abstract}
Reconstructing visual stimulus images is a significant task in neural decoding, and up to now, most studies consider the functional magnetic resonance imaging (fMRI) as the signal source. However, the fMRI-based image reconstruction methods are difficult to widely applied because of the complexity and high cost of the acquisition equipments. Considering the advantages of low cost and easy portability of the electroencephalogram (EEG) acquisition equipments, we propose a novel image reconstruction method based on EEG signals in this paper. Firstly, to satisfy the high recognizability of visual stimulus images in fast switching manner, we build a visual stimuli image dataset, and obtain the EEG dataset by a corresponding EEG signals collection experiment. Secondly, the deep visual representation model (DVRM) consisting of a primary encoder and a subordinate decoder is proposed to reconstruct visual stimuli. The encoder is designed based on the residual-in-residual dense blocks to learn the distribution characteristics between EEG signals and visual stimulus images, while the decoder is designed based on the deep neural network to reconstruct the visual stimulus image from the learned deep visual representation. The DVRM can fit the deep and multiview visual features of human natural state and make the reconstructed images more precise. Finally, we evaluate the DVRM in the quality of the generated images on our EEG dataset. The results show that the DVRM have good performance in the task of learning deep visual representation from EEG signals and generating reconstructed images that are realistic and highly resemble the original images.
\end{abstract}


\section{Introduction}
\IEEEPARstart{N}{eural} decoding is an important way to understand the characteristics of brain visual function\cite{ZHANG2020}. Decoding the visual features of the brain can be widely used in medical rehabilitation\cite{cruz2021self}, human-computer interaction\cite{hekmatmanesh2021review}, entertainment and games\cite{wang2019towards}. Neuroscience research has found that there is a mapping from visual stimuli to brain activity, which takes visual stimuli as input and generates corresponding brain activity\cite{2015Progress, 2019Combination}. Most studies focus on classifying the brain activity, and reconstructing visual stimulus images by the deep generative models\cite{RN2019, FAHIMI2021, 2018Generative, 2022PAN}, such as variational autoencoders (VAEs) and generative adversarial networks (GANs). However, the limited amount of brain activity data generally has problems of low signal--to--noise ratio, extremely high dimensionality and limited spatial resolution\cite{2014AH}. In addition, the inconsistent distribution and representation between the brain activity and the visual images cause great ``domain gap'', and the research on obtaining semantic representation from brain activity is more complex. In a word, the task of reconstructing visual stimulus images from brain activity is still challenging.

Traditionally, the works on visual stimulus images reconstruction are mainly focused on how deep learning can be used to decode and reconstruct visual stimulus images from the functional magnetic resonance imaging (fMRI)\cite{2019Wave2Vec}. For example, Ren \textit{et al.} combined the advantages of adversarial representation learning with knowledge distillation, and proposed a dual-variational autoencoder/generative adversarial network to reconstruct visual stimulus images\cite{2021Reconstructing}. VanRullen \textit{et al.} used a GAN unsupervised procedure to train a VAE neural network, and translated VAE latent codes into reconstructed face images\cite{RN2019}. Du \textit{et al.} used Bayesian deep learning to build a deep generative multiview model\cite{2019Reconstructing}, which translated the visual stimulus images reconstruction into a Bayesian inference problem of missing views in a multiview latent variable model. Next, they use multitask transfer learning of deep neural network (DNN) representations and a matrix-variate Gaussian prior to build a hierarchically structured neural decoding framework, which can reconstruct the perceived natural images with high quality\cite{2022Du}. In addition, they propose a novel doubly semi-supervised multimodal learning (DSML) framework to control the semantics of the imputed modalities accurately, which can recover the missing modality with high visual quality\cite{2021Du}. The above methods of reconstructing visual stimulus images depend on the high spatial resolution and high sensitivity of the fMRI technology. However, considering the fMRI acquisition equipment is bulky, costly and difficult to operate, the above methods are difficult to be widely applied.

Compared to the fMRI acquisition equipment, the equipment of the electroencephalogram (EEG) has the advantages of lower cost, portability and higher temporal resolution\cite{2018An}. Based on the above advantages, the visual stimulus images reconstruction from EEG signals has great promotion significance. However, the EEG signals have a lower signal--to--noise ratio and lower spatial resolution, and the signal processing algorithm must achieve a higher accuracy\cite{2011Convolutional}. Kavasidis \textit{et al.} used an encoder to process the collected signals and extracted features vectors (EEG features) containing category discrimination information\cite{2017deep}. Afterwards, they compared the performance of VAE and GAN in the task of reconstructing the image\cite{2017Brain2Image}. Behnam \textit{et al.} proposed the time-dependent model, and verified the features of the time-dependent model are considered the leading property vector\cite{RN31494}. Zheng \textit{et al.} proposed a combined long short-term
memory--convolutional neural network architecture that extracted the latent features manifold from EEG signals, and employed an improved spectral normalization generative adversarial network to conditionally generate visual stimulus images\cite{2020Decoding}. Seyed \textit{et al.} developed an adaptive neuro-fuzzy inference system topology to obtain the optimal representation of the data\cite{seyed2022retrograde}. The above results strongly demonstrate that the visual relevant representation extracted from EEG signals can effectively generate images semantically consistent with visual stimuli. However, these methods rarely build the mapping relationship between the EEG signals and the visual stimulus images fundamentally, while use powerful generative network models to make up for the lack of the learned deep visual representation from EEG signals.

In this paper, we propose a deep visual representation model (DVRM) to learn deep visual representation from EEG signals and decode the deep visual representation by DNN independent of image distribution features. Firstly, a visual stimuli dataset containing 62 types of the printed letter and number images is designed, and the EEG signals are recorded to build the EEG dataset according to the visual stimuli dataset. Secondly, the DVRM made up of the encoder and decoder is used to learn the deep visual representation and reconstruct the visual stimulus images. On the one hand, the encoder with residual-in-residual dense blocks (RRDB) is the main part of the DVRM. It can adapt to the hierarchical multiview structure of human visual system processing image and learn the deep visual representation from the corresponding EEG signals. On the other hand, the decoder with 7-layer DNN is used to reconstruct images from the learned representation, which will enable the deep visual representation to be decoded directly into images without adding any discriminator of real visual images. Finally, the performance evaluations conducted on the built EEG dataset demonstrate that the DVRM performs well in the task of generating reconstructed images from the learned EEG representation.

The rest of this paper is organized as follows: Section \Rmnum{2} describes the design of experimental paradigm and the construction of dataset; Section \Rmnum{3} presents our proposed DVRM and introduces the whole model in detail; the experiment and results are discussed in Section \Rmnum{4}; Section \Rmnum{5} provides conclusions and an overview of future work.

\section{EEG dataset building based on the designed visual stimuli dataset}
Although the types of letters contained in the Chars74k, NIST and EMNIST datasets are complete and numerous, they all have the same problem when they are used as stimulus images of EEG signals: the recognizability of different characters is low when the images are presented to the subject in a fast switching manner. In this section, we design a visual stimuli dataset which contains the complete printed letter (lowercase, uppercase) and number images. Then, according to the designed visual stimuli dataset, we present a signal acquisition paradigm to collect the EEG signals and build an EEG dataset.

\subsection{Visual stimuli dataset designing}
\begin{figure}[!t]
\centering
\includegraphics[scale=0.54]{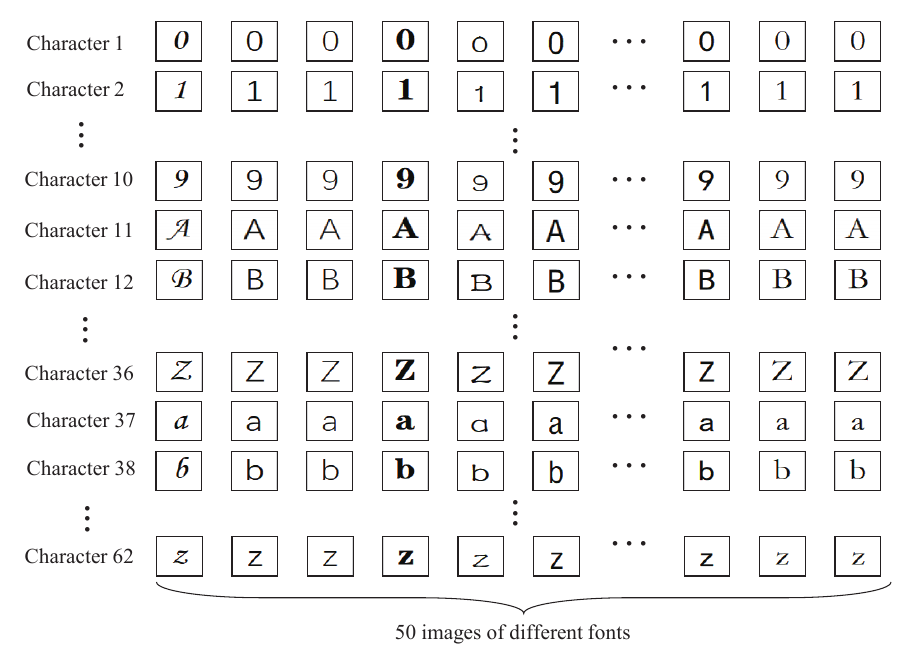}
\caption{The visual stimulus images (62 types of characters, and 50 fonts of each character).}
\label{fig1}
\end{figure}

We collect 260 sets of characters for different fonts in Microsoft Office Professional Plus 2010, and construct their images at a resolution of $28\times 28$. Then, we invite 10 reviewers to evaluate the recognizability of the 260 sets of characters. All reviewers are healthy, have normal vision (no color blindness and color weakness), 22 to 45 years old, and have bachelor's degree or above. We project all the images of each character at the rate of one image per second, and ask the reviewers to record the letters or numbers they see. A score is given according to the proportion of correctly recognised characters (Table \ref{T1}). Considering the interference of fatigue caused by longstanding visual stimulus and the low discrimination of EEG signals induced by similar images, we finally remove the font with the score of 0, 1, 2 and 3, and select 50 fonts with score of 5 to construct the stimulus images. The visual stimulus images contain 62 classes (26 characters: A to Z; 26 characters: a to z; 10 characters: 0 to 9), and each classes contains 50 images (Fig. \ref{fig1}).

\begin{table}[!t]
\begin{center}
\caption{The evaluation of different fonts.}
\label{T1}
\begin{tabular}{ccc}
\hline
Recognition & Scores & Number of fonts \\ \hline
100\%                 & 5      & 154             \\
80\%                  & 4      & 46              \\
60\%                  & 3      & 6               \\
40\%                  & 2      & 13              \\
20\%                  & 1      & 11              \\
0\%                   & 0      & 30              \\ \hline
\end{tabular}
\end{center}
\end{table}

\subsection{EEG Dataset Building}
\begin{figure*}[!t]
\centering
\includegraphics[scale=0.58]{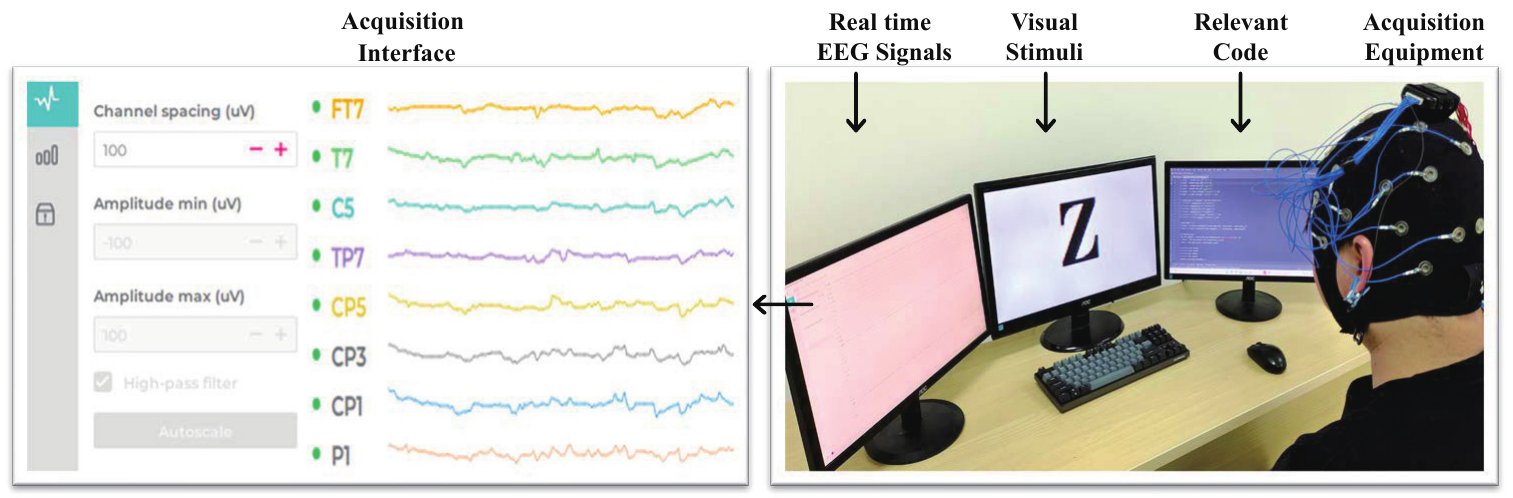}
\caption{The experiment diagram of EEG signals acquisition.}
\label{fig2}
\end{figure*}

The EEG signals are recorded from 4 subjects that are two female and two males, 22 to 26 years old and have uniform educational and cultural background. They have been thoroughly examined to confirm that they are healthy and do not have eye diseases. 4 subjects are trained in signal acquisition for 3 to 4 months before formal signal acquisition begin. The EEG signals are recorded while the visual stimulus images are presented to the subjects. The Fig. \ref{fig2} shows an example of recording EEG signals from a subject. After introducing the purpose of the experiment and relevant precautions to all subjects, they signed the informed consent, and the experiment was approved by the Academic Ethics Committee of Xi 'an University of Science and Technology.

\begin{figure*}[!t]
\centering
\subfloat[]{\includegraphics[width=3in]{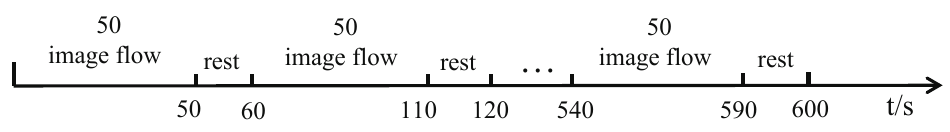}%
\label{fig3a}}
\hfil
\subfloat[]{\includegraphics[width=3in]{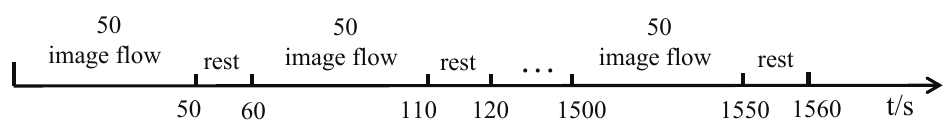}%
\label{fig3b}}
\caption{The experimental paradigm: (a) The experimental paradigm about images of numbers; (b) The experimental paradigm about images of letters.}
\label{fig3}
\end{figure*}

The EMOTIV EPOC Flex is used as the EEG acquisition equipment to collect the EEG signals. This device is a portable 32 channels EEG acquisition device with sampling frequency of 128Hz. In order to remove the EEG noise interference signal, the baseline correction is performed on the recorded signals which are filtered by a 1 to 64 bandpass filter. The time at which each image began to be displayed is taken as time 0, and the 1000ms time before time 0 is taken as the baseline latency range. The baseline correction is achieved by subtracting the average of the previous 1000ms signals from the effective EEG signal. Filter data using Hamming windowed sinc FIR filter. Lower edge of the frequency pass band is 64 Hz, and higher edge is 1 Hz. The experiment is divided into three parts, and each part records the EEG signals generated by the visual stimulus images of numbers, uppercase and lowercase respectively. After one part, the subjects rest for about ten minutes. During the recording session in one part, each type of character images is shown successively in 100s, and the two images are separated by 1s idle periods (1s per image, 1s between and 1 figure for 1 event). Therefore, EEG signals corresponding to each type of character image contain 50 trails. After one type of character, a 10 second rest period is followed when a black image is shown (Fig. \ref{fig3}). The black image is used to ``clear'' the brain activity caused by watching the previous type of character images. During our experiment, the first 100 time steps from each EEG trail are discarded to minimize the disturbance from the previous stimulus image.

\section{Deep Visual Representation Extracting and Image Reconstructing}
Assuming that $D=\{(x_i, y_i)|x_i\in X,y_i\in Y\}_{i=1}^N$ denotes the training set which consists $N$ pairs of samples. $X$ is the set of visual stimulus image, and $Y$ is the corresponding set of EEG data. We introduce the latent space $Z=\{z_i|z_i\in \mathbb{R}^d\}_{i=1}^N$ to describe the deep visual representation, where $d$ stands for the dimension of latent variable $z_i$. The goal of our work is to find the mapping relationship $f:Y\rightarrow X$. In this way, for testing set of EEG data $Y^{*}$, the corresponding visual stimulus image can be generated through $X^{pred}=f(Y^{*})$. Therefore, we first map the valid information between visual stimulus images and brain activity signals into the latent space. This mapping can be expressed as $f_1:(X,Y)\rightarrow Z$. Then, we reconstruct the images based on the mapping $f_2:Z\rightarrow X$.

\begin{figure*}[!t]
\centering
\subfloat[]{\includegraphics[width=6in]{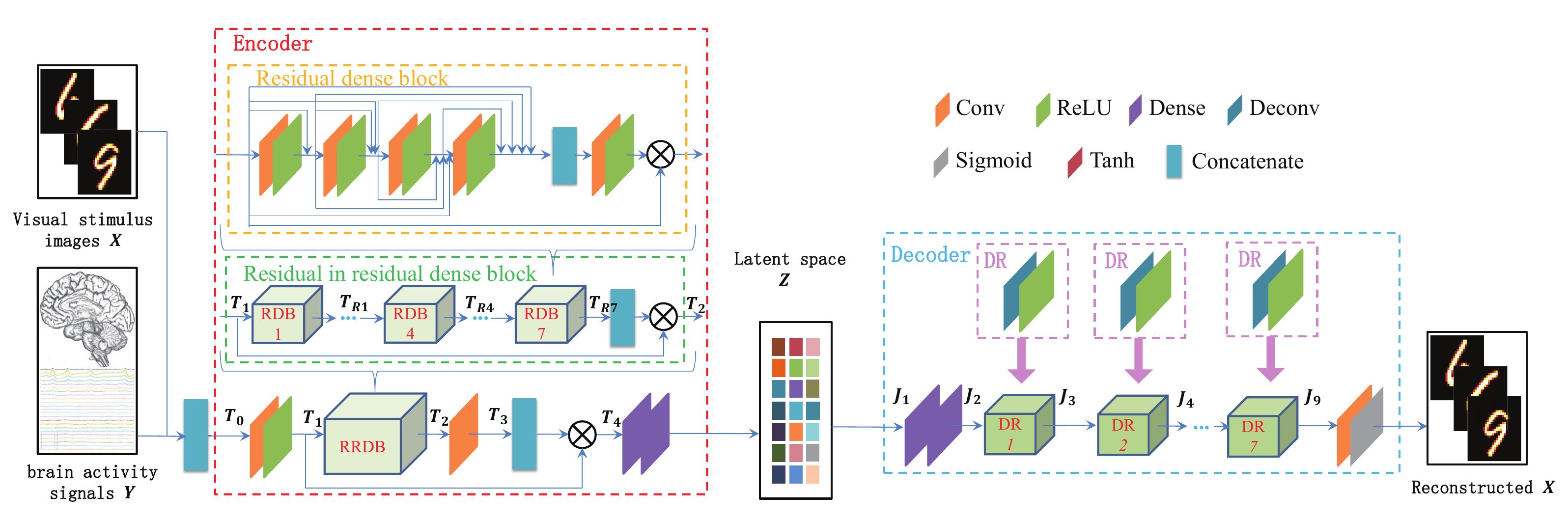}%
\label{fig4a}}
\hfil
\subfloat[]{\includegraphics[width=6in]{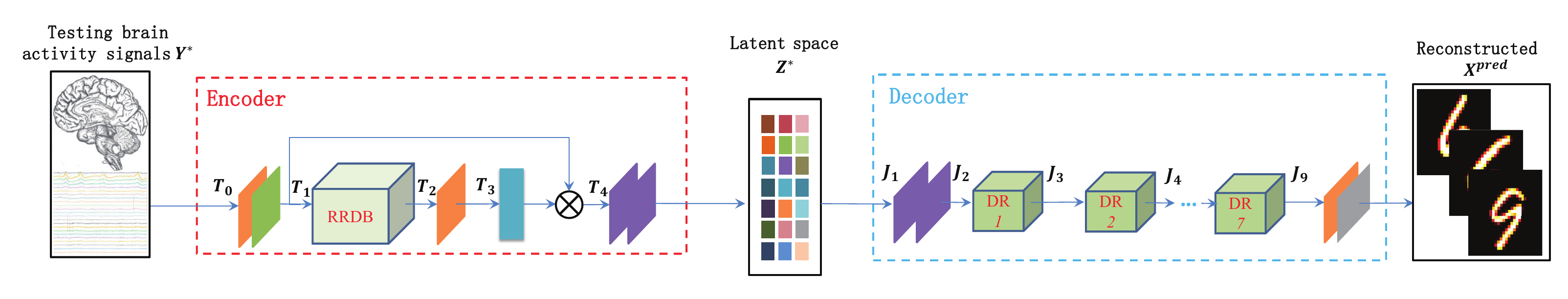}%
\label{fig4b}}
\caption{The framework of DVRM. (a) Training process: $X$ and $Y$ are input to the encoder to obtain $Z$, which is used to reconstruct $X$ through the decoder. (b) Test process: $Y^{*}$ is encoded to $Z^{*}$ by the pretrained encoder. For $Z^{*}$, the pretrained decoder can reconstruct $X^{pred}$. }
\label{fig4}
\end{figure*}

To achieve the above process, we propose the DVRM, and its framework is shown in Fig. \ref{fig4}. The DVRM consists of an encoder ($H_{E}$) and a decoder ($H_{D}$). In the training stage, the encoder based on the RRDB is introduced to obtain the latent space $Z$ from $Y$ and corresponding $X$. The encoder learns the correlation distribution features of EEG signals and corresponding visual stimulus images in this process. For $Z$, the decoder based on the DNN is used to reconstruct visual images (Fig. \ref{fig4a}). In the test stage, we use the pretrained encoder to get the latent space $Z^{*}$ by the set of  brain activity signals $Y^{*}$. The pretrained encoder can extract the visual representation from the EEG signals in this process. For $Z^{*}$, the visual stimulus images can be reconstructed by the decoder (Fig. \ref{fig4b}).

\subsection{Deep Visual Representation Extraction}
In order to learn the visual representation between $X$ and $Y$, we design an encoder with the RRDB to model the mapping $f_1$. The structure diagram of the encoder is shown in Fig. \ref{fig4a}. $X$ and $Y$ are concatenated to obtain the features $T_{0}$, and $T_{0}$ are input into the first convolutional layer to obtain the shallow features $T_{1}$. The features $T_{1}$ are input into the RRDB for deeper features extraction. The RRDB contains $7$ residual dense blocks (RDBs), the output $T_{2}$ of the RRDB can be derived as
\begin{equation}
T_{2}=H_{R D B, 7}\left(H_{R D B, 6}\left(\cdots\left(H_{R D B, 1}\left(T_{1}\right)\right) \cdots\right)\right)\label{eq1}
\end{equation}
where $H_{R D B, d}(d=1, 2,\cdots ,7)$ represents the operation of the $d$-th RDB. Since $T_{2}$ is generated by the $7$-th RDB after making full use of each convolutional layer in the block, it can be regarded as a local feature. Then, the extracted features $T_{2}$ are input to the convolution layer for further feature extraction to get features $T_{3}$. After concatenating the features $T_{3}$ and shallow features $T_{1}$, the global features $T_{4}$ are obtained and then two fully connected layers are used to map $T_{4}$ to a higher dimensional space. Specifically, given $X$ and $Y$, $H_{E}$ produces a distribution over the possible values of $Z$. Inspired by the probabilistic interpretation of latent variables\cite{2014Stochastic}, we define the distribution $p_{\theta}(Z|(X,Y))$ as
\begin{equation}
p_{\theta}(Z|(X,Y))=\prod \limits_{i=1}^N \mathcal{N}(z_i;\mu(x_i,y_i),\textrm{diag}(\sigma^2(x_i,y_i)))\label{eq2}
\end{equation}
where the mean $\mu(x_i,y_i)$ and the diagonal covariance $\sigma^2(x_i,y_i)$ are the output of $H_{E}$ parameterized by $\theta$. We use the Kullback-Leibler (KL) divergence between $p_{\theta}(Z|(X,Y))$ and the prior distribution $p(Z)$ to regularize $p_{\theta}(Z|(X,Y))$. The KL divergence loss $\mathcal{L}_{KL}$ is shown as
\begin{align}
\mathcal{L}_{KL}&=D_{KL}(p_{\theta}(Z|(X,Y))\|p(Z))  \notag\\
&=D_{KL}(\prod \limits_{i=1}^N \mathcal{N}(z_i;\mu(x_i,y_i),\textrm{diag}(\sigma^2(x_i,y_i)))\notag\\
&\qquad\qquad\qquad\qquad \| \prod \limits_{i=1}^N \mathcal{N}(z_i;\mu_z,\textrm{diag}(\sigma_z^2)))\label{eq3}
\end{align}
where the $D_{KL}$ is the KL divergence\cite{kingma2013auto}. We assume that each dimension of the $Z$ should contain the certain semantic information, which makes the model more interpretable. Therefore, the prior distribution $p(Z)$ is assumed to be isotropic Gaussian distributions.
\begin{equation}
p(Z)=\prod \limits_{i=1}^N \mathcal{N}(z_i;0,I)\label{eq4}
\end{equation}
where the isotropic Gaussian distribution with zero-mean and diagonal covariance is consistent with previous works\cite{2014Stochastic,2014Generative}. So Equation (\ref{eq3}) can be expressed as
\begin{align}
\mathcal{L}_{KL}= D_{KL}(\prod \limits_{i=1}^N \mathcal{N}(z_i;\mu(x_i,y_i),\textrm{diag}(\sigma^2(x_i,y_i))) \notag\\
\|\prod \limits_{i=1}^N \mathcal{N}(z_i;0,I))\label{eq5}
\end{align}

\subsection{Visual Stimulus Image Reconstruction}
Given $Z$, we design the decoder based the DNN to reconstruct the corresponding visual stimulus images. Compared with the GAN\cite{RN31438,RN31444}, the DNN network has two advantages in the task of reconstructing visual stimulus images. On the one hand, the DNN architecture can capture the hierarchical visual representation from $Z$, which resembles the visual processing of human. On the other hand, although the capacity of DNN is limited in generating image, the completeness of visual representation can be better embodied by the DNN. So we use the decoder based on the DNN to model the mapping $f_2$. The structure diagram of the decoder is shown in Fig. \ref{fig4a}. The features $J_{1}$ are input into the convolutional layer to get the features $J_{2}$. For the convenience of presentation, we denote the deconvolution layer and the activation layer as DR block. The decoder contains 7 DR blocks. After first convolutional layer in decoder, $J_{2}$ goes through the first to 7-th DR block in turn, the features of dimensionality reduction $J_{9}$ are derived as
\begin{equation}
J_{9} =H_{DR, 7}\left(H_{DR, 6}\left(\cdots\left(H_{DR, 1}\left(J_{2}\right)\right) \cdots\right)\right)\label{eq6}
\end{equation}
where $H_{DR, d}$ can be regarded as the composite function of deconvolution and activation function, which represents the operation of the $d$-th DR. Specifically, we assume the image pixels follow a multivariate Gaussian distribution with zero-mean and diagonal covariance\cite{2016Ladder,2014Semi}. Therefore, the distribution $p_{\varphi}(X|Z)$ can be defined as
\begin{equation}
p_{\varphi}(X|Z)=\prod \limits_{i=1}^N \mathcal{N}(x_i;\mu(z_i),\textrm{diag}(\sigma^2(z_i)))\label{eq7}
\end{equation}
where $\mu(z_i)$ and $\sigma^2(z_i)$ denote the mean and covariance, and they are obtained by the nonlinear transformations of $z_i$. We use the DNN to implement these nonlinear transformations, and take the mean and covariance as the output of $H_{D}$ parameterized by $\varphi$. According to the structural characteristics of DNN, we can define the loss of reconstruction as
\begin{equation}
\mathcal{L}_{rec}=-\mathbb{E}_{p_{\theta}(Z|(X,Y))}[\log p_{\varphi}(X|Z)]\label{eq8}
\end{equation}
For getting the reconstructed $X$ according to the latent space $Z$, we can updata the parameters $\theta$ and $\varphi$ by minimizing the loss $\mathcal{L}$ which can be defined as
\begin{align}
\mathcal{L}&= \mathcal{L}_{rec}+ \mathcal{L}_{KL} =-\mathbb{E}_{p_{\theta}(Z|(X,Y))}[\log p_{\varphi}(X|Z)] \notag\\
&~~~+D_{KL}(\prod \limits_{i=1}^N \mathcal{N}(z_i;\mu(x_i,y_i),\textrm{diag}(\sigma^2(x_i,y_i)))\notag\\
&\qquad\qquad\qquad\qquad\qquad\qquad\quad \|\prod \limits_{i=1}^N \mathcal{N}(z_i;0,I))\label{eq9}
\end{align}

\begin{table*}[!t]
\begin{center}
\caption{The details of character combinations} \label{T2}
\begin{tabular}{cccccc}
\hline
No.       &Combinations                             & Training/Validation/Test images   & Kinds     & Total images      & Specific                                                                                                                                                            \\ \hline
1            &number and number                        & 400/50/50                  & 2      & 500            & ``0-1'',``2-8'',``3-7'',``4-5'',``6-9''                                                                                                                             \\
\multirow{2}{*}{2}           &\multirow{2}{*}{lowercase and lowercase} & \multirow{2}{*}{400/50/50} & \multirow{2}{*}{2}  & \multirow{2}{*}{1300} & \multirow{2}{*}{\begin{tabular}[c]{@{}c@{}}``a-z'',``b-y'',``c-x'',``d-w'',``e-v'',``f-u'',``g-t'',\\ ``h-s'',``i-q'',``j-r'',``k-p'',``l-o'',``m-n''\end{tabular}} \\
                                         &                        &                    &                                                                                                                                                                     \\
\multirow{2}{*}{3}            &\multirow{2}{*}{uppercase and uppercase} & \multirow{2}{*}{400/50/50} & \multirow{2}{*}{2}  & \multirow{2}{*}{1300} & \multirow{2}{*}{\begin{tabular}[c]{@{}c@{}}``A-Z'',``B-Y'',``C-X'',``D-W'',``E-V'',``F-U'',``G-T'',\\ ``H-S'',``I-R'',``J-Q'',``K-P'',``L-N'',``M-O''\end{tabular}} \\
                                         &                        &                    &                                                                                                                                                                     \\
4           &number and lowercase                     & 400/50/50                  & 2        & 400        & ``0-h'',``2-g'',``e-4'',``f-3''                                                                                                                                     \\
5           &number and uppercase                     & 400/50/50                  & 2        & 400          & ``7-X'',``S-8'',``6-D'',``K-2''                                                                                                                                     \\
6           &lowercase and uppercase                  & 400/50/50                  & 2          & 400        & ``g-D'',``p-Q'',``R-W'',``H-a''                                                                                                                                     \\
7           &``BRAINS''                               & 240/30/30                 & 6          & 300        & ``B-R-A-I-N-S''                                                                                                                                                     \\ \hline
\end{tabular}
\end{center}
\end{table*}

\section{Experimental results and analysis}
This section discusses the feasibility of DVRM-based visual stimulus images reconstruction. We provide results and discussion from two aspects :1) We analyze the performance of the Encoder module and evaluate how it learns to extract potential EEG visual representations; 2) We used four evaluation methods to evaluate the quality of reconstructed images.
\subsection{Classification of EEG deep representation}
We implement the DVRM in the Python 3.7 with the Tensorflow 2.1.0. The computer is with the GPU of the NVIDIA GeForce GTX 1080 Ti, the CPU of the AMD Ryzen 5 2600 processor and the RAM of 32GB. We divided the EEG dataset (26 characters: A to Z; 26 characters: a to z; 10 characters: 0 to 9) three parts: training set, validation set, and test set, which are divided as 80\%, 10\%, and 10\%, respectively.

We choose two common deep neural networks, AlexNet and residual networks (ResNet), to compare with RRDB in DVRM. In the training stage, the learning rate is initially set to 0.001, and the Encoder is trained using the Adam gradient descent, with the batch size of 20. The state parameters of the model and hyperparameters of training were fine-tuned on the validation set. We use the classical k--nearest--neighbour classifier to classify the representation vector about characters of different categories. The comparison results of the three encoders are shown in the Table \ref{TADD1}. The data in Table \ref{TADD1} are the average classification accuracy of the four subjects on the validation set and the test set.

\begin{table}[!t]
\begin{center}
\setlength{\tabcolsep}{7mm}
\caption{Maximum validation accuracy and associated test accuracy for three different encoders.} \label{TADD1}
\begin{tabular}{lll}
\hline
Encoder & Validation & Test    \\ \hline
AlexNet & 83.24\%    & 81.98\% \\
ResNet  & 72.71\%    & 69.31\% \\
RRDB    & 86.56\%    & 83.21\% \\ \hline
\end{tabular}
\end{center}
\end{table}

\begin{figure}[!t]
\centering
\includegraphics[scale=0.3]{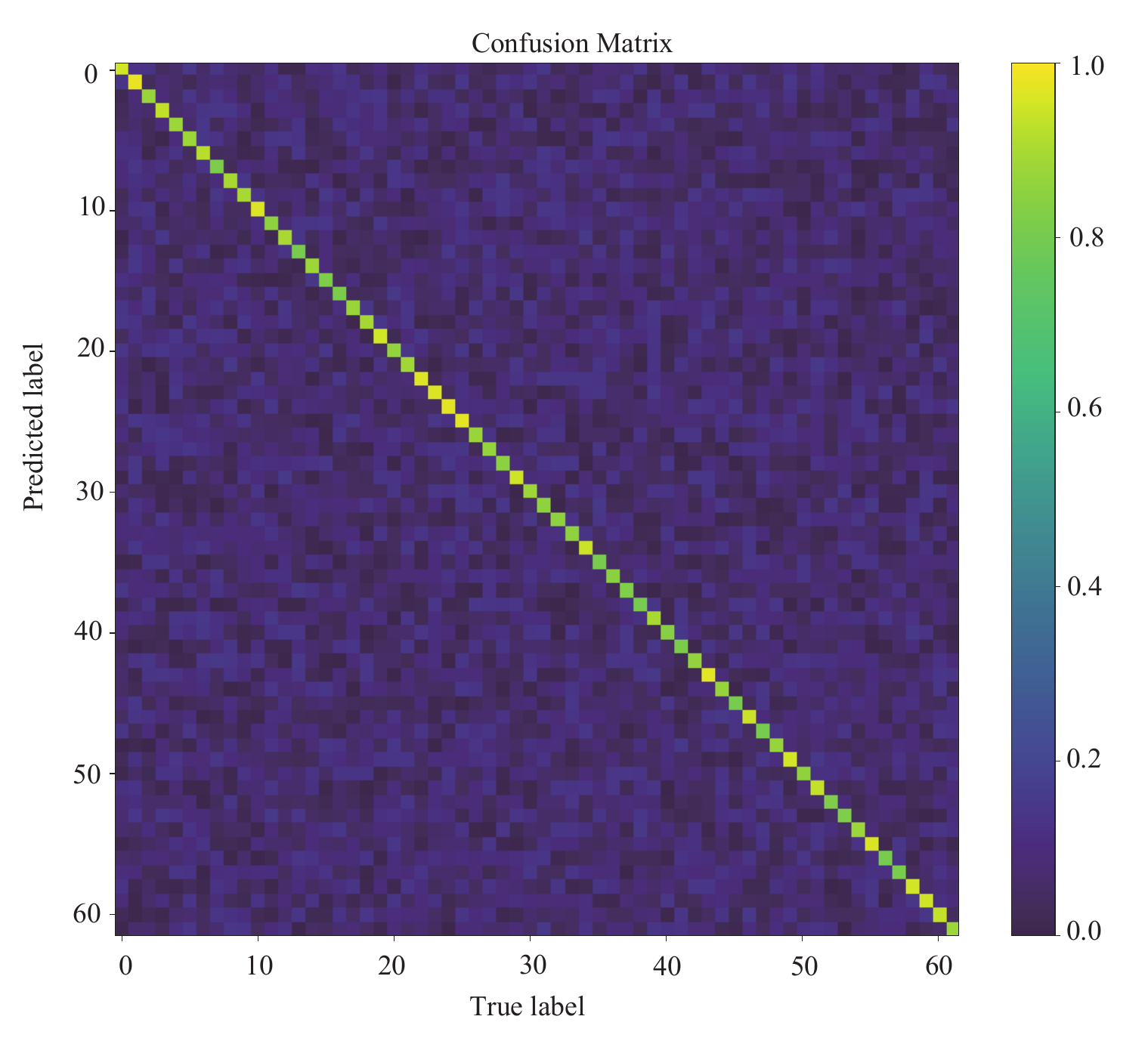}
\caption{The confusion matrixes of our RRDB-based encoder.}\label{figadd}
\end{figure}

In the encoder, the accuracy of our RRDB is 1.23\% higher than the AlexNet, and 13.9\% higher than the ResNet. The residual structure in the ResNet is an essential element in the RRDB. In RRDB, the residual structure can be used to extract not only global but also local features. Therefore RRDB as an encoder is significantly better than ResNet. Although AlexNet works much better than ResNet, it still falls short when compared to RRDB. The confusion matrix of our RRDB-based encoder is provided in Fig. \ref{figadd}. As seen, a satisfactory classification result can be reached in most of the categories. These experimental results seem to confirm that our idea is right. By introducing RRDB, we are able to simultaneously harness the ability of residual structure to extract global and local features and the capability of dense structure to obtain all the layered features.

\subsection{Image Reconstruction and Evaluation Metrics}

In the training stage, we extract two classes of character images (40 images in per character) from uppercase images, lowercase images and number images randomly, and build 6 kinds of combinations for image training respectively. Specially, to further validate the model performance, we also build character combination which includes the uppercase (``B'', ``R'', ``A'', ``I'', ``N'', ``S'') to train the DVRM. The combinations are shown in Table \ref{T2}. We train the DVRM on 80\% of the combined images and corresponding EEG signals. The input of the DVRM is $N\times 28\times 28$ character images and the corresponding $N\times 32\times 135$ EEG signals, where $N$ is the amount of data of data samples, $28\times 28$ is the number of pixels of the image, 32 is the number of channels of the EEG signal, and 135 is the time step of an EEG signals. The data is first fed into the RRDB of the encoder. For the RRDB, it consists of convolutional layers with the $kernel\_size=2$, $strides=1$ and $padding=same$. Furtherly, the shared latent features obtained by the encoder are further fed into the decoder composed of deconvolution layers to reconstruct the visual stimulus images.

\begin{figure}[!t]
\centering
\includegraphics[scale=0.25]{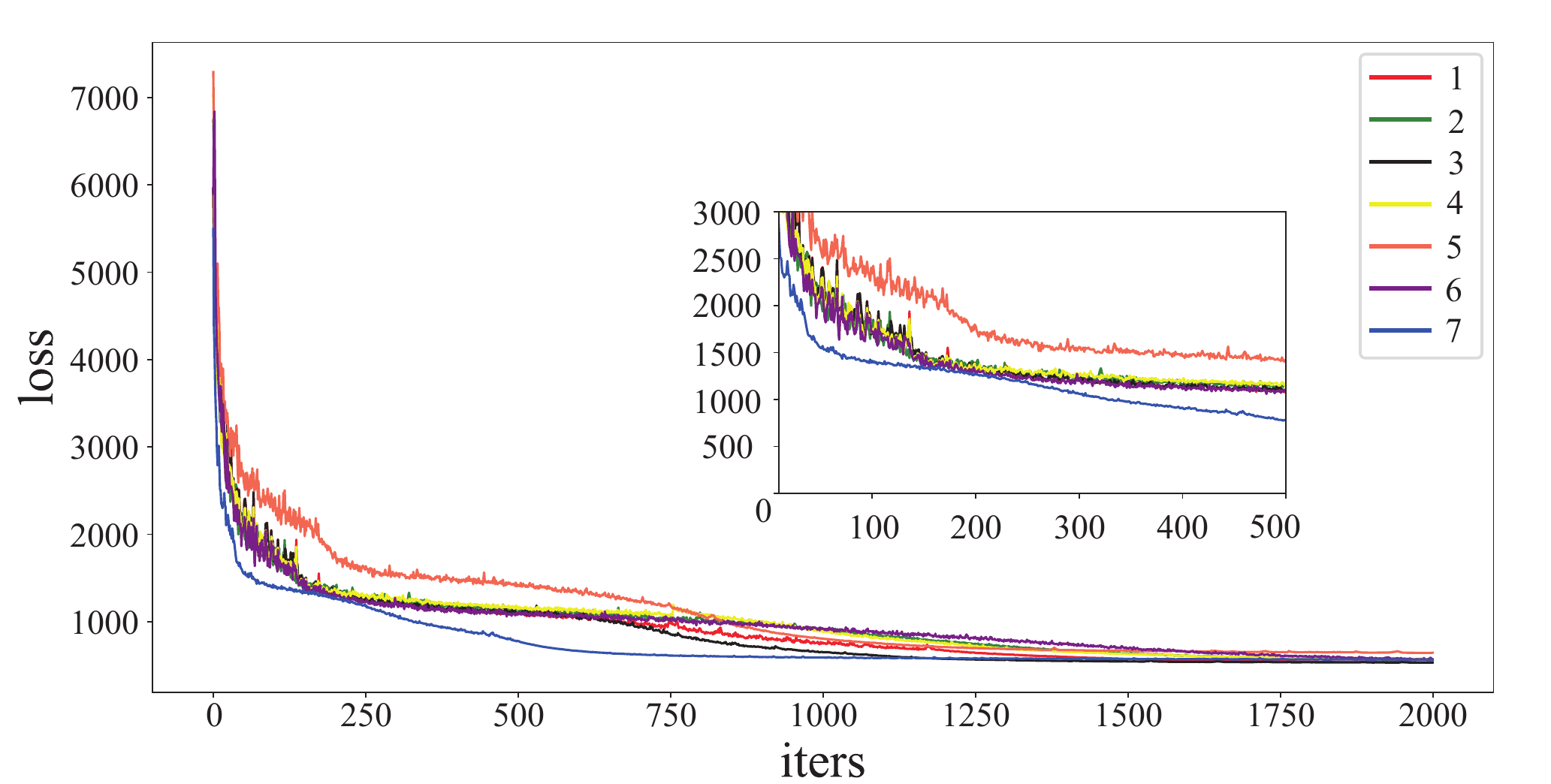}
\caption{Loss value with the number of iterations in training stage.}\label{fig5}
\end{figure}

In addition, we adopt Adam algorithm to optimize the DVRM\cite{kingma2014adam}. The parameters of Adam algorithm, learning rate $lr$, the first decay rate $\beta_1$ and the second decay rate $\beta_2$ are initialized as $2\times 10^{-5}$, $0.9$ and $0.999$. The partial relation between the loss value of different character combinations training and the number of iterations is shown in Fig. \ref{fig5}, the loss can stabilize at 2000 iterations. So we use 2000 iterations of back-propagation to constitute an epoch. The total number of parameters for the whole model training are 7258155. All the parameters are used to training model. In the testing stage, we use EEG signals as input to generate visual stimulus images.

In order to quantify the performance of reconstructed images by the DVRM, we use the following metrics to evaluate the similarity between the visual stimulus image and the reconstructed image.

\emph{1) Pearson's correlation coefficient (PCC)\cite{benesty2009pearson}.} The PCC varies from $-$1 to 1. The larger coefficient value, the stronger linear correlation between two sets of data. A positive coefficient value represents a positive correlation, and a negative value represents a negative correlation. Here we tile the 2D matrix data of the reconstructed image into the 1D vector to calculate the PCC between the two images.

\emph{2) Structural Similarity (SSIM)\cite{SSIM}.} Due to the consideration of image texture characteristics, SSIM has a good correlation with human visual perception. It has a value between $[0,1]$, and the higher value, the better performance. The experiment uses utils library of Python to calculate the SSIM metric between two images.

\emph{3) Peak signal--to--noise ratio (PSNR)\cite{korhonen2012peak}.} The PSNR is often used as a measurement method of signals reconstruction quality in image compression and other fields. It is the mean square error between the original image and the processed image relative to $(2^n-1)^2$.

\emph{4) Mean square error (MSE)\cite{Hyndman2006another}.} The MSE method first needs to calculate the mean square value of the pixel difference between the original image and the reconstructed image, and then determine the degree of distortion of the distorted image by the magnitude of the mean square value.

\subsection{Reconstruction Results and Analysis}
\begin{figure*}[!t]
\centering
\subfloat[]{\includegraphics[width=8cm]{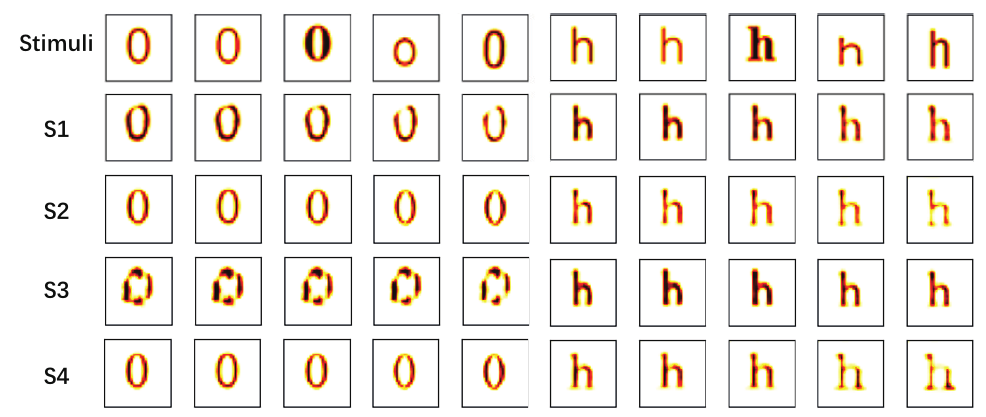}%
\label{fig6a}}
\hfil
\subfloat[]{\includegraphics[width=8cm]{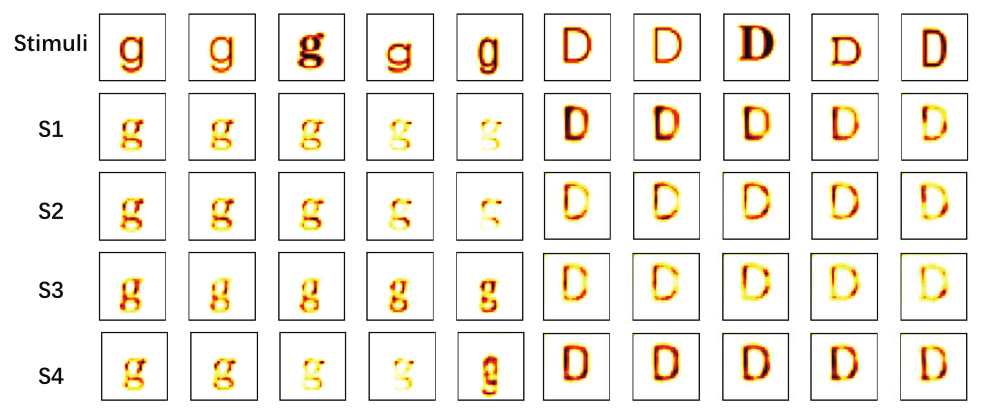}%
\label{fig6b}}
\hfil
\subfloat[]{\includegraphics[width=8cm]{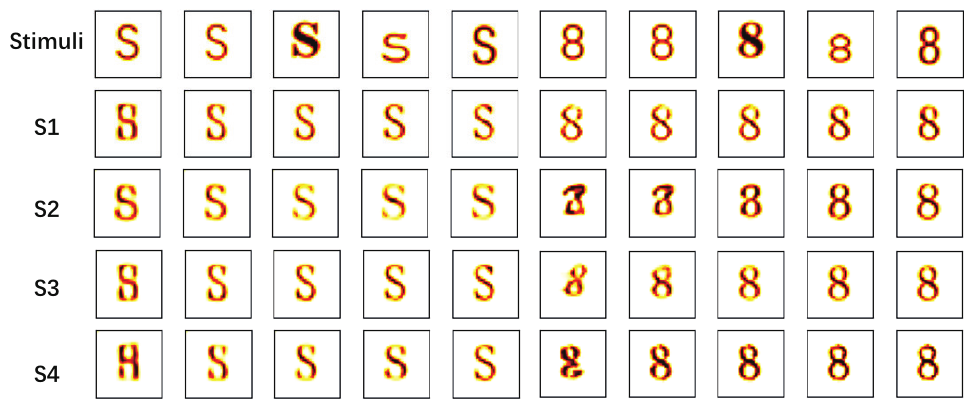}%
\label{fig6c}}
\hfil
\subfloat[]{\includegraphics[width=8cm]{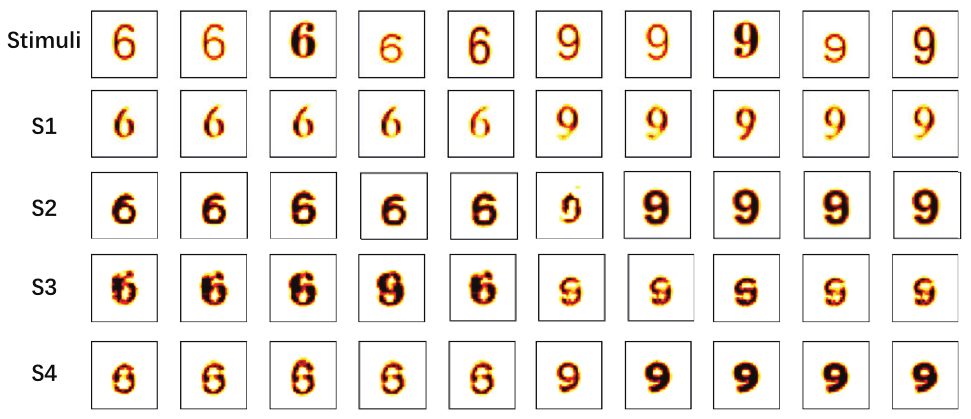}%
\label{fig6d}}
\caption{The Partially reconstructed images: (a) ``0-h''; (b) ``g-D''; (c) ``S-8''; (d) ``6-9''.}
\label{fig6}
\centering
\includegraphics[scale=0.8]{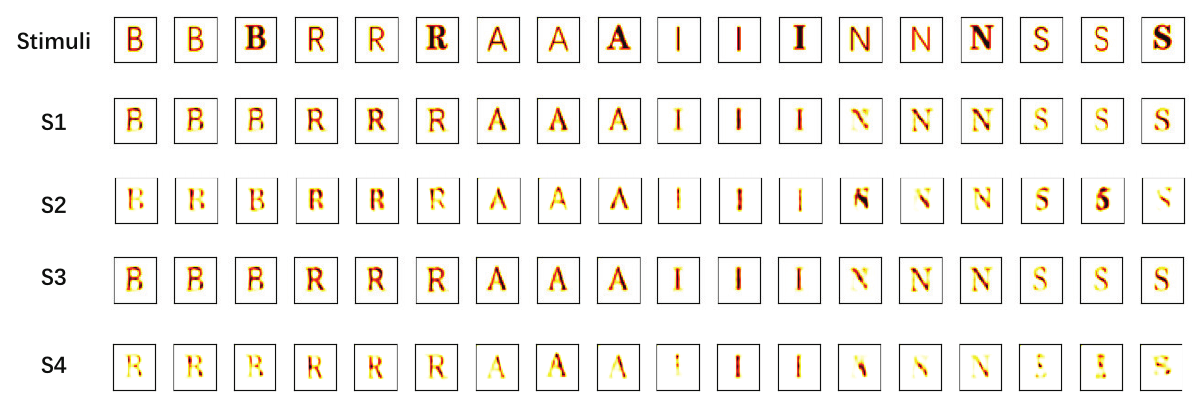}
\caption{Partially reconstructed images of ``BRAINS''.}
\label{fig7}
\end{figure*}

The partial reconstructed images generated by four subjects (S1, S2, S3 and S4) are shown in Figs. 7(a)--(d) and \ref{fig7}, which show that we successfully reconstruct the image from EEG signals evoked by visual stimuli through the DVRM. The complete reconstruction results are shown in the appendix Tables \ref{T5}--\ref{T8}. All the values in these tables are the average of the results obtained after repeating 10 tests. To describe the reconstruction effectiveness between different subjects more generally, we calculate the average values of four subjects about each character combination, and the results are shown in Table \ref{T3}.

\begin{table*}[!t]
\begin{center}
\caption{The reconstruction results of the DVRM (mean$\pm$ std)} \label{T3}
\begin{tabular}{cccccc}
\hline
\multirow{2}{*}{No.}    & \multirow{2}{*}{Stimuli} & \multicolumn{4}{c}{Metrics}                                                                                                         \\ \cline{3-6}
                         &  & PCC                             & SSIM                            & PSNR                             & MSE                             \\ \hline
1   &Number and number            & 0.544$\pm$.035                     & 0.435$\pm$0.038                     & 14.705$\pm$1.359                     & 0.029$\pm$0.006                     \\
2   &Lowercase and lowercase      & 0.555$\pm$0.079                     & 0.458$\pm$0.062                     & 16.246$\pm$0.848                     & 0.025$\pm$0.006                     \\
3   &Uppercase and Uppercase      & 0.514$\pm$0.075                     & 0.378$\pm$0.061                     & 14.633$\pm$1.032                     & 0.034$\pm$0.007                     \\
4    &Number and lowercase         & 0.508$\pm$0.028                     & 0.437$\pm$0.041                     & 14.948$\pm$0.851                    & 0.031$\pm$0.006                     \\
5   &Number and uppercase         & 0.515$\pm$0.029                     & 0.450$\pm$0.029                     & 14.430$\pm$1.511                     & 0.029$\pm$0.007                     \\
6   &Uppercase and lowercase      & 0.495$\pm$0.038                     & 0.437$\pm$0.030                     & 14.132$\pm$1.368                     & 0.033$\pm$0.007                     \\
\textbf{-}    &\textbf{Average}                      & \textbf{0.522$\pm$0.047}                     & \textbf{0.433$\pm$0.044}                     & \textbf{14.849$\pm$1.162}                     & \textbf{0.030$\pm$0.007} \\ \hline
7   &``BRAINS''                   &0.512$\pm$0.058                        & 0.407$\pm$0.031                     &13.779$\pm$1.478                     & 0.038$\pm$0.007 \\ \hline
\end{tabular}
\end{center}
\end{table*}

From Table \ref{T3}, it can be found that the values of PCC, SSIM, PSNR and MSE are up to 0.555, 0.458, 16.246 and 0.025 for the combination of ``lowercase and lowercase''. The reconstructed ``lowercase and lowercase'' images performs better than the other combinations. In addition, comparing with the combinations of different types character, it can be found that combinations of same type character are relatively better. For the reconstruction results of all characters combinations, the average values of PCC, SSIM, PSNR and MSE are up to 0.522, 0.433, 14.849 and 0.030. Specially, for the combination of ``BRAINS'' in Table \ref{T3}, the values of PCC, SSIM, PSNR and MSE are up to 0.512, 0.407, 13.779 and 0.038.

To show the effectiveness of our proposed model, we compare the evaluation metrics of reconstructed images (``BRAINS'' and ``combinations of two types character'') with those of \cite{2019Reconstructing}. In \cite{2019Reconstructing}, the values of PCC and SSIM are up to 0.813 and 0.651 for the combination ``6-9'', and are up to 0.502 and 0.360 for the combination``BRAINS''. For our model, the average values of PCC and SSIM are up to 0.522 and 0.433 for the combinations of `` two types character'', and the values of PCC and SSIM are up to 0.512 and 0.407 for the combination``BRAINS''. The results show that performance of our reconstructed ``BRAINS'' images is relatively better. Although the performance of the reconstructed ``combinations of two types character'' images is slightly good in \cite{2019Reconstructing}, our reconstructed images still have high identifiability in terms of visual experience. Therefore, we conclude that all of the reconstructed images have good performance, and the proposed DVRM can effectively reconstruct highly similar visual stimulus images.

\begin{rem}
 It should be noted that, we reconstruct image from EEG signals based on the proposed DVRM, while \cite{2019Reconstructing} from fMRI signals based on  bayesian deep multiview learning method. The reason why we compare with \cite{2019Reconstructing} is that, the adopted evaluation metrics (``PCC'', ``SSIM'')  are the same with ours.
\end{rem}

\subsection{Further Discussions}
\begin{itemize}
\item[1)]\emph{Reconstructing the Same Character with Different Appearances:} In the visual stimulus images, the appearance of some letters varies greatly from different fonts. For example, the letters `g' also can be written as another form. But it can be observed in Fig. 7(b) that the sample `g' can still be reconstructed in the form of most fonts when the presented image changes a great deal. It confirm that the DVRM can extract the deep visual representation that mainly contains the semantic information of the image content and reconstruct the image with same semantic information of presented image according to the visual representation.

\item[2)]\emph{Analyzing the Distinctions in the reconstructed performance:} The reconstructed results are distinguishing between different character combinations. For the given latent variables of the visual stimulus image and the corresponding brain activity, we ideally assume that it is isotropic Gaussian distributions so that the each dimension of the visual representation contains the certain semantic information\cite{2014Stochastic,2014Generative}. Realistically, the each dimension of the visual representation is interrelated rather than independent. Therefore, taking into account the interpretability of the representation and the complexity of the calculation is an improvement direction.
\end{itemize}

\section{Conclusion}
In this paper, we propose a DVRM for learning the deep visual representation and reconstructing the visual stimulus images. On the one hand, we design the encoder based on the RRDB to learn the deep visual representation between the visual stimulus images and the corresponding brain activity. On the other hand, we adopt 7-layer DNN to generate the reconstructed images. Due to lack of relevant dataset, we design the visual stimuli dataset and build the EEG dataset according to the collected EEG signals. We evaluate the DVRM through the quality of the generated images on the built EEG dataset. The reconstructed results show that we successfully reconstruct the image from EEG signals evoked by visual stimuli through the DVRM, and reconstructed image is highly similar to the original image.



\bibliographystyle{IEEEtran}
\bibliography{Reference}

\begin{onecolumn}
\appendix[The complete reconstruction results for all subjects and character combinations]

\setcounter{table}{0}   
\renewcommand{\thetable}{A\arabic{table}}

\begin{table}[h!]
\setlength{\abovecaptionskip}{-0.27cm}
\setlength{\belowcaptionskip}{-0.2cm}
\begin{center}
\caption{The reconstruction results of number images} \label{T5}
\resizebox{\textwidth}{!}{
\begin{tabular}{ccccccccccccccccc}
\hline
\multirow{2}{*}{Stimuli} & \multicolumn{4}{c}{S1} & \multicolumn{4}{c}{S2} & \multicolumn{4}{c}{S3} & \multicolumn{4}{c}{S4} \\ \cmidrule(l){2-5}\cmidrule(l){6-9}\cmidrule(l){10-13}\cmidrule(l){14-17}
                         & PCC   & SSIM  & PSNR   & MSE   & PCC   & SSIM  & PSNR   & MSE   & PCC   & SSIM  & PSNR   & MSE   & PCC   & SSIM  & PSNR   & MSE   \\ \hline
0-1                        & 0.546                   & 0.449                    & 12.977                   & 0.023                   & 0.572                   & 0.480                    & 12.677                   & 0.026                   & 0.522                   & 0.390                    & 12.040                   & 0.036                   & 0.548                   & 0.434                    & 12.702                   & 0.020                   \\
2-8                        & 0.511                   & 0.446                    & 15.548                   & 0.035                   & 0.492                   & 0.457                    & 15.242                   & 0.029                   & 0.512                   & 0.395                    & 16.075                   & 0.038                   & 0.513                   & 0.391                    & 15.628                   & 0.022                   \\
3-7                        & 0.510                   & 0.472                    & 14.630                   & 0.026                   & 0.553                   & 0.309                    & 13.220                   & 0.051                   & 0.546                   & 0.455                    & 13.643                   & 0.038                   & 0.529                   & 0.450                    & 13.435                   & 0.035                   \\
4-5                         & 0.502                   & 0.442                    & 15.901                   & 0.027                   & 0.482                   & 0.413                    & 15.619                   & 0.030                   & 0.481                   & 0.434                    & 15.651                   & 0.030                   & 0.564                   & 0.501                    & 15.791                   & 0.030                   \\
6-9                        & 0.496                   & 0.459                    & 14.903                   & 0.034                   & 0.587                   & 0.322                    & 13.140                   & 0.043                   & 0.546                   & 0.436                    & 13.542                   & 0.029                   & 0.525                   & 0.481                    & 13.441                   & 0.035                   \\\hline
\end{tabular}}
\end{center}
\end{table}

\begin{table}[h!]
\setlength{\abovecaptionskip}{-0.27cm}
\setlength{\belowcaptionskip}{-0.2cm}
\begin{center}
\caption{The reconstruction results of lowercase images} \label{T6}
\resizebox{\textwidth}{!}{
\begin{tabular}{ccccccccccccccccc}
\hline
\multirow{2}{*}{Stimuli} & \multicolumn{4}{c}{S1} & \multicolumn{4}{c}{S2} & \multicolumn{4}{c}{S3} & \multicolumn{4}{c}{S4} \\ \cmidrule(l){2-5}\cmidrule(l){6-9}\cmidrule(l){10-13}\cmidrule(l){14-17}
                         & PCC   & SSIM  & PSNR   & MSE   & PCC   & SSIM  & PSNR   & MSE   & PCC   & SSIM  & PSNR   & MSE   & PCC   & SSIM  & PSNR   & MSE   \\ \hline
a-z                        & 0.498 & 0.406 & 15.500 & 0.029 & 0.551 & 0.443 & 15.232 & 0.032 & 0.533 & 0.460 & 15.675 & 0.029 & 0.532 & 0.434 & 15.244 & 0.031 \\
b-y                        & 0.640 & 0.539 & 16.383 & 0.024 & 0.624 & 0.520 & 16.092 & 0.025 & 0.505 & 0.413 & 15.366 & 0.031 & 0.545 & 0.425 & 15.952 & 0.027 \\
c-x                        & 0.652 & 0.508 & 17.943 & 0.018 & 0.675 & 0.529 & 18.184 & 0.017 & 0.632 & 0.453 & 17.540 & 0.018 & 0.577 & 0.432 & 16.956 & 0.021 \\
d-w                        & 0.485 & 0.423 & 16.021 & 0.021 & 0.402 & 0.346 & 15.231 & 0.025 & 0.488 & 0.421 & 15.873 & 0.021 & 0.486 & 0.403 & 15.109 & 0.020 \\
e-v                        & 0.535 & 0.462 & 16.451 & 0.023 & 0.421 & 0.387 & 15.544 & 0.028 & 0.529 & 0.437 & 16.277 & 0.024 & 0.509 & 0.423 & 15.988 & 0.021 \\
f-u                        & 0.447 & 0.392 & 16.131 & 0.027 & 0.385 & 0.357 & 15.840 & 0.030 & 0.535 & 0.444 & 16.667 & 0.024 & 0.467 & 0.408 & 16.002 & 0.020 \\
g-t                        & 0.630 & 0.497 & 16.529 & 0.023 & 0.656 & 0.536 & 16.422 & 0.023 & 0.643 & 0.511 & 16.198 & 0.025 & 0.619 & 0.512 & 16.350 & 0.025 \\
h-s                        & 0.524 & 0.407 & 15.824 & 0.029 & 0.533 & 0.413 & 15.835 & 0.030 & 0.559 & 0.426 & 16.282 & 0.028 & 0.465 & 0.434 & 15.675 & 0.021 \\
i-q                        & 0.551 & 0.484 & 16.708 & 0.025 & 0.578 & 0.494 & 17.659 & 0.023 & 0.631 & 0.544 & 18.270 & 0.021 & 0.645 & 0.504 & 16.612 & 0.024 \\
j-r                        & 0.521 & 0.527 & 17.096 & 0.021 & 0.486 & 0.487 & 16.643 & 0.024 & 0.532 & 0.525 & 17.103 & 0.021 & 0.439 & 0.382 & 14.904 & 0.033 \\
k-p                        & 0.509 & 0.375 & 15.907 & 0.032 & 0.562 & 0.430 & 15.894 & 0.031 & 0.514 & 0.360 & 15.287 & 0.033 & 0.449 & 0.363 & 14.766 & 0.035 \\
l-o                        & 0.607 & 0.544 & 17.732 & 0.021 & 0.587 & 0.535 & 17.716 & 0.022 & 0.670 & 0.570 & 18.046 & 0.019 & 0.583 & 0.506 & 16.056 & 0.027 \\
m-n                        & 0.611 & 0.466 & 15.630 & 0.029 & 0.560 & 0.432 & 14.234 & 0.039 & 0.739 & 0.592 & 16.834 & 0.023 & 0.657 & 0.489 & 15.584 & 0.029 \\ \hline
\end{tabular}}
\end{center}
\end{table}

\begin{table}[h!]
\setlength{\abovecaptionskip}{-0.27cm}
\setlength{\belowcaptionskip}{-0.2cm}
\begin{center}
\caption{The reconstruction results of uppercase images} \label{T7}
\resizebox{\textwidth}{!}{
\begin{tabular}{ccccccccccccccccc}
\hline
\multirow{2}{*}{Stimuli} & \multicolumn{4}{c}{S1}                                                                                 & \multicolumn{4}{c}{S2}                                                                                 & \multicolumn{4}{c}{S3}                                                                                 & \multicolumn{4}{c}{S4}                                                                                 \\  \cmidrule(l){2-5}\cmidrule(l){6-9}\cmidrule(l){10-13}\cmidrule(l){14-17}
                         & PCC                       & SSIM                      & PSNR                       & MSE                       & PCC                       & SSIM                      & PSNR                       & MSE                       & PCC                       & SSIM                      & PSNR                       & MSE                       & PCC                       & SSIM                      & PSNR                       & MSE                       \\ \hline
A-Z                        & 0.556                     & 0.371                     & 14.915                     & 0.034                     & 0.618                     & 0.437                     & 15.789                     & 0.029                     & 0.530                     & 0.355                     & 14.667                     & 0.036                     & 0.532                     & 0.323                     & 14.622                     & 0.031                     \\
B-Y                        & 0.617                     & 0.421                     & 15.079                     & 0.024                     & 0.473                     & 0.368                     & 14.042                     & 0.025                     & 0.505                     & 0.355                     & 15.366                     & 0.031                     & 0.545                     & 0.425                     & 15.952                     & 0.027                     \\
C-X                        & 0.562                     & 0.421                     & 15.691                     & 0.030                     & 0.531                     & 0.428                     & 15.058                     & 0.035                     & 0.555                     & 0.442                     & 15.326                     & 0.033                     & 0.620                     & 0.459                     & 16.058                     & 0.029                     \\
D-W                        & 0.464                     & 0.230                     & 12.871                     & 0.054                     & 0.498                     & 0.298                     & 13.206                     & 0.053                     & 0.435                     & 0.297                     & 12.751                     & 0.048                     & 0.475                     & 0.342                     & 15.234                     & 0.032                     \\
E-V                        & 0.465                     & 0.330                     & 14.582                     & 0.037                     & 0.397                     & 0.270                     & 13.945                     & 0.042                     & 0.542                     & 0.381                     & 14.921                     & 0.035                     & 0.399                     & 0.357                     & 14.404                     & 0.037                     \\
F-U                        & 0.500                     & 0.463                     & 13.854                     & 0.035                     & 0.586                     & 0.478                     & 14.055                     & 0.033                     & 0.409                     & 0.389                     & 14.142                     & 0.032                     & 0.452                     & 0.329                     & 14.804                     & 0.035                     \\
G-T                        & 0.475                     & 0.355                     & 14.648                     & 0.031                     & 0.457                     & 0.341                     & 15.009                     & 0.039                     & 0.581                     & 0.442                     & 13.438                     & 0.040                     & 0.415                     & 0.314                     & 14.632                     & 0.036                     \\
H-S                        & 0.610                     & 0.396                     & 15.292                     & 0.035                     & 0.572                     & 0.365                     & 14.639                     & 0.039                     & 0.552                     & 0.350                     & 14.759                     & 0.040                     & 0.622                     & 0.502                     & 16.432                     & 0.023                     \\
I-R                        & 0.587                     & 0.425                     & 15.630                     & 0.033                     & 0.611                     & 0.452                     & 16.652                     & 0.032                     & 0.611                     & 0.452                     & 16.652                     & 0.032                     & 0.643                     & 0.506                     & 16.230                     & 0.023                     \\
J-Q                        & 0.446                     & 0.314                     & 14.121                     & 0.034                     & 0.467                     & 0.359                     & 14.406                     & 0.037                     & 0.593                     & 0.397                     & 13.684                     & 0.039                     & 0.435                     & 0.334                     & 14.172                     & 0.037                     \\
K-P                        & 0.592                     & 0.379                     & 15.057                     & 0.035                     & 0.405                     & 0.344                     & 13.754                     & 0.034                     & 0.544                     & 0.332                     & 14.333                     & 0.039                     & 0.434                     & 0.352                     & 14.037                     & 0.031                     \\
L-N                        & 0.471                     & 0.322                     & 15.014                     & 0.037                     & 0.455                     & 0.332                     & 14.627                     & 0.039                     & 0.431                     & 0.292                     & 15.096                     & 0.039                     & 0.423                     & 0.336                     & 14.455                     & 0.033                     \\
M-O                        & 0.444                     & 0.330                     & 12.155                     & 0.033                     & 0.503                     & 0.380                     & 12.494                     & 0.029                     & 0.518                     & 0.410                     & 12.490                     & 0.028                     & 0.599                     & 0.470                     & 14.784                     & 0.024                     \\ \hline
\end{tabular}}
\end{center}
\end{table}

\begin{table}[h!]
\setlength{\abovecaptionskip}{-0.27cm}
\setlength{\belowcaptionskip}{-0.2cm}
\begin{center}
\caption{The reconstruction results of random images} \label{T8}
\resizebox{\textwidth}{!}{
\begin{tabular}{ccccccccccccccccc}
\hline
\multirow{2}{*}{Stimuli} & \multicolumn{4}{c}{S1} & \multicolumn{4}{c}{S2} & \multicolumn{4}{c}{S3} & \multicolumn{4}{c}{S4} \\ \cmidrule(l){2-5}\cmidrule(l){6-9}\cmidrule(l){10-13}\cmidrule(l){14-17}
                         & PCC   & SSIM  & PSNR   & MSE   & PCC   & SSIM  & PSNR   & MSE   & PCC   & SSIM  & PSNR   & MSE   & PCC   & SSIM  & PSNR   & MSE   \\ \hline
0-h                                          & 0.537                   & 0.474                    & 14.125                   & 0.030                   & 0.510                   & 0.415                    & 14.274                   & 0.039                   & 0.502                   & 0.399                    & 14.045                   & 0.032                   & 0.506                   & 0.469                    & 14.710                   & 0.030                   \\
2-g                                          & 0.470                   & 0.500                    & 16.013                   & 0.025                   & 0.561                   & 0.485                    & 15.840                   & 0.035                   & 0.528                   & 0.487                    & 15.928                   & 0.022                   & 0.505                   & 0.442                    & 16.443                   & 0.027                   \\
e-4                                          & 0.445                   & 0.436                    & 14.921                   & 0.042                   & 0.513                   & 0.425                    & 14.149                   & 0.042                   & 0.503                   & 0.435                    & 14.512                   & 0.039                   & 0.494                   & 0.369                    & 14.116                   & 0.024                   \\
f-3                                          & 0.520                   & 0.376                    & 14.807                   & 0.028                   & 0.477                   & 0.429                    & 14.740                   & 0.023                   & 0.515                   & 0.441                    & 15.495                   & 0.034                   & 0.540                   & 0.409                    & 15.051                   & 0.028                   \\
7-X                                          & 0.490                   & 0.467                    & 12.336                   & 0.037                   & 0.457                   & 0.446                    & 12.776                   & 0.032                   & 0.495                   & 0.425                    & 12.059                   & 0.035                   & 0.516                   & 0.499                    & 12.606                   & 0.027                   \\
S-8                                          & 0.503                   & 0.425                    & 14.953                   & 0.037                   & 0.504                   & 0.400                    & 14.180                   & 0.023                   & 0.526                   & 0.471                    & 14.982                   & 0.023                   & 0.561                   & 0.457                    & 15.000                   & 0.036                   \\
6-D                                          & 0.501                   & 0.463                    & 14.470                   & 0.024                   & 0.547                   & 0.466                    & 14.866                   & 0.023                   & 0.536                   & 0.422                    & 14.148                   & 0.034                   & 0.487                   & 0.426                    & 14.342                   & 0.036                   \\
K-2                                          & 0.537                   & 0.430                    & 16.126                   & 0.035                   & 0.549                   & 0.485                    & 15.780                   & 0.023                   & 0.495                   & 0.464                    & 16.328                   & 0.021                   & 0.535                   & 0.456                    & 15.933                   & 0.019                   \\
g-D                                          & 0.470                   & 0.440                    & 14.141                   & 0.042                   & 0.490                   & 0.431                    & 14.198                   & 0.040                   & 0.516                   & 0.433                    & 14.504                   & 0.025                   & 0.532                   & 0.406                    & 14.749                   & 0.043                   \\
p-Q                                          & 0.535                   & 0.401                    & 14.890                   & 0.027                   & 0.471                   & 0.417                    & 15.445                   & 0.029                   & 0.539                   & 0.444                    & 15.409                   & 0.036                   & 0.535                   & 0.399                    & 14.832                   & 0.026                   \\
R-W                                          & 0.420                   & 0.464                    & 11.876                   & 0.042                   & 0.452                   & 0.406                    & 12.318                   & 0.040                   & 0.476                   & 0.482                    & 12.497                   & 0.036                   & 0.500                   & 0.478                    & 11.990                   & 0.030                   \\
H-a                                          & 0.484                   & 0.466                    & 14.979                   & 0.021                   & 0.551                   & 0.478                    & 14.303                   & 0.030                   & 0.481                   & 0.442                    & 14.944                   & 0.031                   & 0.468                   & 0.402                    & 15.040                   & 0.030                   \\ \hline
\end{tabular}}
\end{center}
\end{table}
\end{onecolumn}

\end{document}